\documentclass[runningheads]{llncs}
\usepackage[T1]{fontenc}
\usepackage[lofdepth,lotdepth]{subfig}
\usepackage{graphicx}
\usepackage{float}
\usepackage{tabularx}
\usepackage{booktabs}
\usepackage{ragged2e}
\usepackage{longtable}
\usepackage{multirow}
\usepackage{svg}
\usepackage{color, colortbl}
\usepackage{pdfpages}
\usepackage{makecell} 
\PassOptionsToPackage{table,xcdraw, dvipsnames}{xcolor}
\usepackage{xcolor}
\usepackage{inconsolata}
\usepackage{enumitem}

\usepackage{tikz}
\definecolor{cloudwhite}{cmyk}{0,0,0,0.025}  % color
\definecolor{engineering}{rgb}{0.56,0.18, 0.11}
\definecolor{azulvibrante}{RGB}{0, 57, 95}
\definecolor{laranjaintenso}{RGB}{109, 42, 12}

\usepackage{listings}
\usepackage{todonotes}
\presetkeys%
    {todonotes}%
    {inline,backgroundcolor=yellow}{}

\usepackage[hidelinks,bookmarks,bookmarksopen,bookmarksdepth=2]{hyperref}

\usepackage{color}

\begin{document}
\title{Refactoring Towards Microservices:\\ Preparing the Ground for Service Extraction}
\titlerunning{Refactoring Towards Microservices}
\author{
    Rita Peixoto\inst{1} \and
    Filipe F. Correia\inst{1} \and
    Thatiane Rosa\inst{2} \and \\
    Eduardo Guerra\inst{3} \and
    Alfredo Goldman\inst{4}
}
\authorrunning{R. Peixoto et al.}

\institute{
    INESC TEC, Faculty of Engineering, University of Porto, Portugal \\
    \email{\{rita,filipe.correia\}@fe.up.pt} \and
    IFTO, Brazil\\
    \email{thatiane@ifto.edu.br} \and
    UNIBZ, Italy\\
    \email{eduardo.guerra@unibz.it} \and
    IME/USP, Brazil\\
    \email{gold@ime.usp.br}
}

\maketitle
% typeset the header of the contribution
%
\begin{abstract}
As organizations increasingly transition from monolithic systems to microservices, they aim to achieve higher availability, automatic scaling, simplified infrastructure management, enhanced collaboration, and streamlined deployments. However, this migration process remains largely manual and labour-intensive. While existing literature offers various strategies for decomposing monoliths, these approaches primarily focus on architecture-level guidance, often overlooking the code-level challenges and dependencies that developers must address during the migration. This article introduces a catalogue of seven refactorings specifically designed to support the transition to a microservices architecture with a focus on handling dependencies. The catalogue provides developers with a systematic guide that consolidates refactorings identified in the literature and addresses the critical gap in systematizing the process at the code level. By offering a structured, step-by-step approach, this work simplifies the migration process and lays the groundwork for its potential automation, empowering developers to implement these changes efficiently and effectively.

\end{abstract}
\section{Introduction}\label{introduction}
As systems evolve, they inevitably face new challenges that expose the limitations and trade-offs of prior architectural decisions~\cite{Cervantes2016DesigningApproach,Martin2018CleanArchitecture,Richards2020Fundamentals}. A growing trend in response to these challenges is the adoption of microservices architectures, driven by the increasing demands placed on systems, like independent scaling of components, the expansion of codebases and of the teams that work on them, and the increasing complexity of business requirements~\cite{zuber}. This shift, while promising significant benefits, inherently necessitates a code migration process that is both intricate and resource-intensive.

Migrating to a microservices architecture promises numerous advantages, including enhanced flexibility, scalability, and accelerated development cycles. However, the migration process itself is far from straightforward. A successful migration requires careful planning, a deep understanding of the existing system and business aspects, and a clear understanding of the potential impact on both the codebase and the broader system architecture~\cite{newman_monolith}.

In essence, transitioning from a monolithic system to a microservices architecture involves transforming a tightly coupled system into a collection of small, independent, and loosely coupled services. The complexity of this transformation is primarily dictated by the level of code entanglement and dependency within the existing monolithic system~\cite{fowler_howtobreak}.

Despite the growing body of research on microservices migration~\cite{balalaie_microservices_2018,fritzsch_monolith_2019,kalia_mono2micro_2020,freitas_refactoring_2021,correia_identification_2022}, much of it focuses on isolated aspects of the process, particularly the identification of service boundaries, often referred to as microservices candidates. While valuable, such guidance typically concentrates on architectural goals rather than providing actionable, step-by-step assistance to developers. Additionally, existing systematizations in the field tend to focus on patterns at a high level rather than addressing the granular refactoring tasks developers must execute to achieve these patterns.

Architectural refactoring is inherently a complex endeavour that significantly impacts the structure of a system. Despite the existing research, the limited practical guidance in the current literature means much of the migration process relies heavily on developer intuition rather than systematic, well-defined methodologies~\cite{abid202030}. 

Recognizing this gap, this article aims to address the need for a structured approach by presenting a catalogue of refactorings that handle dependencies between systems and facilitate microservices adoption. It is intended for software developers, architects, and technical leads who are directly involved in the migration of monolithic systems to microservices architectures and are interested in systematic approaches to software refactoring and architectural evolution. 

This work tries to look beyond identifying what a good service boundary might be. By focusing on actionable, code-level refactorings, our goal is to bridge the gap between high-level architectural strategies and the practical challenges faced by developers during the migration process. It breaks down large-scale refactorings into sequences of smaller, manageable steps, enabling teams to address dependencies systematically and effectivelly ``preparing the ground'' for isolating a new service. As explained by Fowler~\cite{Fowler99}, large changes are often composed of a series of smaller transformations that preserve behavior and can be applied incrementally. This step-by-step strategy is consistent with the \textsc{Strangler Fig} pattern~\cite{fowler_stranglerfig,yoder_strangler_2022}, which promotes gradually replacing parts of a legacy system while maintaining system stability. This catalogue intends to provide a structured and practical resource that can guide developers through the migration process, reducing risks and making large-scale changes more feasible.

In this article, we use the terms \textit{service} and \textit{microservice} interchangeably, referring to components that exhibit the desirable characteristics of microservices-based architectures. In some cases, these terms may also refer to \textit{components intended to become microservices} after extraction.

In the following sections, we outline the methodology employed to develop this catalogue, detail its contents, and highlight seven refactorings specifically focused on handling dependencies between systems. Next, we present some related work, and finally, we present our final considerations. This work aspires to bridge the gap between high-level architectural strategies and the hands-on refactoring efforts required to transition from monolithic to microservices architectures effectively.

\section{Methodology}\label{methodology}
We built the catalogue presented in this article through three main steps: (i) a review of scientific literature, (ii) analysis of technical and grey literature, and (iii) an industry survey. Our literature review selected 88 papers using predefined inclusion/exclusion criteria and keywords related to microservices migration and refactoring (see replication package\footnote{\scriptsize{Replication package available at: \url{https://github.com/RitaPeixoto/Migration-of-Monoliths-to-Microservices-Survey\_replication\_package}}}). Searches were mainly conducted via Google Scholar. The selected papers were categorized by topics such as service decomposition, large-scale refactoring, migration automation, and related challenges. Most focused on technical aspects; few addressed practitioners' views.

We also drew on two technical references: Newman's book~\cite{newman_monolith} and the foundational article by Lewis and Fowler~\cite{microservices_fowler}. Grey literature~\cite{sevenHard,tolerant_reader,monoliths1,monoliths2}—including blogs, reports, and articles—was used to incorporate practical perspectives into the catalogue.

To gather practitioner insights, we conducted a survey between 2023-01-03 and 2023-06-15, which received 66 responses. The survey aimed to understand how practitioners conduct microservices migrations, the tools they use, and the motivations behind their choices. The main findings from this survey, as well as all supporting materials, are available in our replication package.

Therefore, we started composing the catalogue by identifying the higher-scale refactorings reported in the literature review. From there, we identified the smaller-scale refactorings that they are composed of. To do this, we tried to understand what is needed in each situation to solve the present issue and then break it into smaller steps to make it more feasible.

The process of decomposing higher-scale refactorings into smaller-scale ones involved analyzing what was needed in each situation to resolve existing issues and breaking the process into feasible steps. We refer to one common type of small-scale refactoring as ``breaking dependencies'', whose goal is to break functional dependencies between services. This can mean changing a local dependency to a remote dependency or changing a local dependency to be a local dependency still, but that facilitates the transition to a remote dependency in the future.

Our analysis of Newman’s patterns~\cite{newman_monolith} and survey data suggests that many practices represent intermediate steps rather than complete transitions to microservices. While we do not address specific quality attributes in each refactoring, we assume mechanisms like performance optimization, data consistency, and fault tolerance, and robust testing practices are essential for building are essential for a fully functional microservices system. In particular, automated testing across unit, integration, and end-to-end levels is critical to ensure that independently deployed services behave reliably and evolve safely over time.

\section{A Refactorings Catalog for Handling Dependencies}\label{breaking_dependencies}

To separate the system into microservices, we need to identify the dependencies between the clusters of classes that will make our intended microservices and ``break'' the dependencies between those clusters so that the microservices can function independently.

We can consider multiple types of dependencies. This section catalogues refactorings for evolving code in such a way that dependencies are taken into consideration and evolved accordingly. It is common to find these dependency types together, so some refactorings described below may link to others and form a sequence of refactorings to solve the dependencies between microservices.

As we describe in Section~\ref{introduction}), we focused on describing the refactorings to handle the functional dependencies. We were particularly interested in the most easily actionable refactorings, which became the core focus of our catalogue. These refactorings were identified and refined through the analysis steps described in Section~\ref{methodology}, which involved reviewing academic literature, technical books, and grey literature. Table~\ref{tab:refactorings_references} shows the seven refactorings included in our catalogue and the main references we used. These sources helped identify the common transformation needs and express them as concrete refactorings that preserve the system's behavior.

\begin{longtable}{lcc}
\caption{Refactorings references}
\\\hline
     \textbf{Refactoring Name} & \textbf{References} \\ \hline 
\endfirsthead

\multicolumn{3}{c}%
{{ \tablename\ \thetable{} -- continued from previous page}} \\
\hline \multicolumn{1}{l}{\textbf{Refactoring Name}} & \multicolumn{1}{c}{\textbf{References}} \\ \hline 
\endhead

\hline \multicolumn{3}{r}{{Continued on next page}} \\ \hline
\endfoot

\endlastfoot                                                                          
    Replace Method Call with Service Call~~~ &  \cite{balalaie_microservices_2018}, \cite{siroky_monolith_2021}, \cite{pinto_refactoring_2019} \\
    Move Foreign-key Relationship to Code & \cite{freitas_refactoring_2021}, \cite{newman_monolith}, \cite{pinto_refactoring_2019}\\
    Replicate Data Across Microservices & \cite{fowler_event}\\
    Split Database Across Microservices & \cite{siroky_monolith_2021}, \cite{newman_monolith}\\
    Create Data Transfer Object & \cite{freitas_refactoring_2021}\\
    Break Data Type Dependency & \cite{freitas_refactoring_2021}\\
    Shared code isolation & - \\
    \bottomrule
     \label{tab:refactorings_references}
\end{longtable}

In the next few sections we present our catalog with its different refactorings. We follow a format inspired by the one proposed by Fowler and Beck~\cite{Fowler99}, with additional sections, as follows: 

\begin{itemize}
    \item \textbf{Name:} A concise and descriptive name for the refactoring.
    \item \textbf{Context and motivation:} A description of a set of observed conditions in context, and the rationale for applying the refactoring.
    \item \textbf{Example:} A practical scenario illustrating the state of the system before the refactoring is applied, highlighting the dependencies or issues to be addressed.
    \item \textbf{Strategy:} A high-level explanation of the approach taken to implement the refactoring.
    \item \textbf{Benefits:} A summary of the advantages gained by applying the refactoring, highlighting improvements to the system.
    \item \textbf{Challenges:} A summary of potential difficulties, risks, or trade-offs involved in applying the refactoring.
    \item \textbf{Mechanics:} A detailed, step-by-step guide for applying the refactoring.
    \item \textbf{Example of application:} A practical scenario illustrating the state of the system after the refactoring is applied, showing how the dependencies or issues have been resolved.
\end{itemize}

To make our catalogue easier to understand and apply in practice, we developed the examples using technologies widely adopted in microservices development. All the examples in the catalogue use Java\footnote{Java.com. Available at: \url{https://www.java.com/en/} (Accessed: Jun. 12 2025).} as the programming language. We also use RESTful HTTP for communication between services and Apache Kafka\footnote{``Apache kafka'', Apache Kafka, \url{https://kafka.apache.org/} (Accessed: Jun. 21, 2025).} for event-driven messaging and asynchronous processing.

To illustrate the examples, we consider a hypothetical monolithic system designed for managing order processing and inventory tracking. This system ensures seamless coordination between placing an order and updating the corresponding inventory levels.
Suppose the system contains the following components:
\begin{itemize}
    \item A set of classes responsible for managing orders, which could later be extracted into a microservice called \emph{OrderManagement}.
    \item A set of classes responsible for managing inventory, which could later be extracted into a microservice called \emph{InventoryManagement}.
    \item  A set of classes responsible for managing customer data, which could later be extracted into a microservice called \emph{CustomerManagement}.
\end{itemize}
This system represents a typical scenario in which tightly coupled components, such as order management and inventory services, need to be refactored into independent microservices while maintaining data consistency and seamless communication.

\subsection{Replace Method Call with Service Call}\label{changeCalldoc}

\subsubsection*{Context and motivation}When splitting a monolith into microservices, it is common to encounter code dependencies in the form of direct method invocations between components. These tight couplings make it difficult to extract and deploy services independently. In particular, the presence of local method calls prevents the involved classes from being moved to a separate service, limiting modularity and hindering architectural evolution.

\subsubsection*{Example} Consider the above-mentioned system that manages order processing and inventory tracking, which was initially implemented as a monolith.

A scenario illustrating the need for this refactoring is as follows: \emph{the OrderProcessor class belonging to the OrderManagement domain makes a direct local method call to updateInventory, which resides in the InventoryService class under the InventoryManagement domain.}

Because this interaction is implemented as a local method invocation, it assumes both classes exist within the same runtime and memory space. This dependency means:
\begin{itemize}
    \item Extracting \textit{OrderProcessor} into a separate microservice would break its ability to call \textit{InventoryService} directly.
    \item The communication between order and inventory logic is hidden within internal method calls, rather than exposed through well-defined interfaces or protocols.
    \item \textit{InventoryService} cannot simply be moved to a different service without redesigning how \textit{OrderProcessor} interacts with it.
\end{itemize}

To enable microservice extraction, this local call is replaced with a network call to a new API endpoint within the same service, so that, when a service is extracted in the future, the two components can evolve independently. This is the essence of the refactoring need: transforming the local method call dependency into one based on a network protocol.

\subsubsection*{Strategy}
To decouple these components in preparation for service extraction, we can replace the local method call with explicit service-to-service communication that reflects the intended interaction between future microservices. Such calls should reflect the intended communication between the future services and be implemented using an appropriate protocol, synchronous or asynchronous, depending on the requirements, goals, and constraints.

\subsubsection*{Benefits} This refactoring favors:
\begin{itemize}
    \item Independent service evolution: once local calls are replaced with service interfaces, teams can evolve each service independently without breaking others.
    \item Improved local testability and maintainability: the decoupling of components allows each to be tested in isolation.
    \item Technological reuse: by abstracting communication, services can expose reusable APIs or events, enabling other services to consume them without duplicating logic.
\end{itemize}

\subsubsection*{Challenges} At the same time, this approach introduces important challenges:
\begin{itemize}
    \item Network dependency: local calls are fast and reliable, while remote calls introduce network uncertainty.
    \item Increased latency is a direct consequence of replacing direct method calls with remote ones.
    \item Integration testing difficulties as testing interactions across services becomes more complex once communication is externalized.
\end{itemize}

\subsubsection*{Mechanics}
\begin{enumerate}[itemsep=1ex]
    \item Decide the communication strategy (synchronous or asynchronous).
    \begin{enumerate}
        \item \textbf{Synchronous strategy}\footnote{Occurs when a caller sends a request to a provider service and waits for a response before continuing execution.} should be used
when an immediate response is required, often to ensure data consistency. This form of communication is typically implemented using RPC-style calls, such as gRPC or RESTful HTTP APIs.
\begin{enumerate}
    \item \emph{Benefits}: Can provide low-latency responses, ensures strong consistency, has relatively low implementation complexity, and simplifies the handling of transactional data.
    \item \emph{Consequences}: However, it can negatively affect scalability, availability, performance, fault tolerance, and overall resilience. It also introduces tighter coupling between services. As a result, services depending on synchronous communication may become bottlenecks if the provider service is slow or unavailable, potentially leading to cascading failures. 
    \end{enumerate}
    \item \textbf{Asynchronous strategy}\footnote{Occurs when a caller sends a request to a provider service and continues execution without waiting for an immediate response. This is typically implemented via asynchronous RPC or messaging using a publisher/subscriber model, where services publish messages to a broker, and subscribers consume and process them when available.} is preferable when an immediate response is not required and when eventual consistency is acceptable. Useful when a service call triggers downstream calls, allowing the caller to remain unblocked while waiting for the entire chain of operations to complete. Typically implemented using Event-Driven Architecture (EDA) with message brokers using protocols such as Kafka, RabbitMQ (AMQP), or Mosquitto (MQTT).
    \begin{enumerate}
        \item \emph{Benefits}: This communication style enhances scalability and fault tolerance, improves system availability by decoupling callers from provider failures, and overall system resilience. It also reduces coupling between services, enabling greater flexibility and independence in their evolution.
         \item \emph{Consequences}: However, asynchronous calls are not suitable when real-time (instantaneous) responses or strict data consistency at the time of action are required. In this model, consistency is eventual, which may not meet all business needs. Additionally, asynchronous interactions are more complex to implement and demand robust mechanisms for monitoring, error handling, and message delivery guarantees.
    \end{enumerate}

    \end{enumerate}
    \item Make the initial configurations for the chosen strategy.
    \begin{enumerate}
        \item \underline{Synchronous} (e.g, REST or gRPC)
        \begin{enumerate}
            \item Store the necessary information (\textit{e.g.}, URL) to make remote calls to the target microservice.
        \end{enumerate}
        \item \underline{Asynchronous} (EDA with Kafka, RabbitMQ/AMQP, Mosquitto/MQTT): using strategies like Event Sourcing or some form of asynchronous RPC.
        \begin{enumerate}
            \item Set up a message broker or event bus.
            \item Create a topic or channel for message exchange.
        \end{enumerate}
    \end{enumerate}
    \item Configure the caller microservice - change the local method calls to be remote calls to the provider.
    \begin{enumerate}
        \item \underline{Synchronous}
        \begin{enumerate}
            \item Create an interface with the declaration of the identified methods to call.
            \item Create a class that implements that interface and makes the remote service calls, a Request Class.
        \end{enumerate}
        \item \underline{Asynchronous}
        \begin{enumerate}
            \item The caller subscribes to the created topic/channel: implement subscription logic to receive messages using the chosen broker/protocol (Kafka, AMQP, MQTT).
        \end{enumerate}
    \end{enumerate}
    \item Configure the provider microservice.
    \begin{enumerate}[partopsep=3ex]
        \item \underline{Synchronous} - Provide an API to respond to caller requests:
        \begin{enumerate}
            \item Create a class defining resource paths and request handling.
            \item Add methods to perform the actions requested by the caller.
        \end{enumerate}
        \item \underline{Asynchronous}
        \begin{enumerate}
            \item The provider publishes messages to the topic/channel: implement publishing logic to push messages to the topic using the chosen broker/protocol (Kafka, AMQP, MQTT).
        \end{enumerate}
    \end{enumerate}
\end{enumerate}

\textbf{Important:} Ensure fault tolerance is implemented in the communication strategy to ensure system reliability. Use mechanisms such as retries, circuit breakers, or chaos testing, as detailed in the related documentation (see item C.7.2)~\footnote{\scriptsize{Complementary references available in the related documentation: \url{https://github.com/RitaPeixoto/Migration-of-Monoliths-to-Microservices-Survey_replication_package/blob/main/catalogue_of_refactorings.pdf}}}.

\subsubsection*{Example of application}
In Listings~\ref{initial_service_call_doc} and ~\ref{initial_service_call_doc2}, we can see the current code of the monolith where the \textit{OrderProcessor} class from candidate \textit{OrderManagement} microservice makes a local method call to the method \textit{updateInventory} of the \textit{InventoryService} class of the proposed \textit{InventoryManagement} microservice.

\begin{lstlisting}[language=Java, label=initial_service_call_doc, caption={[Local method call dependency in the Monolith]The \textit{OrderProcessor} class.}]
// Candidate for the OrderManagement microservice
public class OrderProcessor {

    private final InventoryService inventoryService;

    public OrderProcessor(InventoryService inventoryService) {
        this.inventoryService = inventoryService;
    }

    public void processOrder(Order order) {
        // Process the order
        // ...

        // Update inventory after order is processed
        inventoryService.updateInventory(order);
    }
}
\end{lstlisting}
\begin{lstlisting}[language=Java, label=initial_service_call_doc2, caption={[Local method call dependency in the Monolith]The \textit{InventoryService} class.}]
// Candidate for the InventoryManagement microservice
public class InventoryService {

    private final InventoryManager inventoryManager;

    public InventoryService(InventoryManager inventoryManager) {
        this.inventoryManager = inventoryManager;
    }

    public void updateInventory(Order order) {
        // Delegate inventory update logic
        inventoryManager.updateInventory(order);
    }
}
\end{lstlisting}

As in the future these two classes will belong to different microservices, we need to refactor this dependency. 
For that, the method calls should become remote service calls. 

\subsubsection*{Synchronous solution} An example for a synchronous solution using RESTful HTTP is shown in Listings~\ref{final_service_call_sync_doc} and ~\ref{final_service_call_sync_doc2}. 

\begin{lstlisting}[language=Java, label=final_service_call_sync_doc, caption={[Changing local call to synchronous service call in the \textit{OrderManagement} microservice]Synchronous solution for the \emph{OrderManagement} microservice.}]
// Candidate for the OrderManagement microservice
public interface InventoryService {
    void updateInventory(Order order);
}

@Service
public class RemoteInventoryService implements InventoryService {

    private final RestTemplate restTemplate;
    private final String inventoryServiceUrl;

    public RemoteInventoryService(RestTemplate restTemplate,
                                  @Value("${inventory.service.url}") String inventoryServiceUrl) {
        this.restTemplate = restTemplate;
        this.inventoryServiceUrl = inventoryServiceUrl;
    }

    @Override
    public void updateInventory(Order order) {
        String endpoint = inventoryServiceUrl + "/api/inventory/update";
        restTemplate.postForObject(endpoint, order, Void.class);
    }
}

public class OrderProcessor {

    private final InventoryService inventoryService;

    public OrderProcessor(InventoryService inventoryService) {
        this.inventoryService = inventoryService;
    }

    public void processOrder(Order order) {
        // Business logic for processing the order
        // ...
        // Synchronous call to update inventory
        inventoryService.updateInventory(order);
    }
}
\end{lstlisting}
\begin{lstlisting}[language=Java, label=final_service_call_sync_doc2, caption={[Changing local call to synchronous service call in the \textit{OrderManagement} microservice]Synchronous solution for the \emph{InventoryManagement} microservice.}]
// Candidate for the InventoryManagement microservice
@RestController
@RequestMapping("/api/inventory")
public class InventoryController {

    @PostMapping("/update")
    public ResponseEntity<Void> updateInventory(@RequestBody Order order) {
        // Logic to update inventory based on the order
        // ...
        return ResponseEntity.ok().build();
    }
}
\end{lstlisting}

In the \textit{OrderManagement} microservice, we created the \textit{InventoryService} interface that defines the contract for inventory updates (\textit{updateInventory} method). Then, we created the implementation of this interface, the \textit{RemoteInventoryService} class that makes RESTful calls to the \emph{InventoryManagement} microservice. Finally, the \textit{OrderProcessor} class remains the same, although now it depends on the \textit{InventoryService} interface to update inventory after processing an order.

In the \textit{InventoryManagement} microservice, we created the \textit{InventoryController} class that handles the HTTP POST request for updating the inventory. The \textit{updateInventory} method updates the inventory based on the received order.

This refactoring changes the direct local method call dependency to a synchronous RESTful API call, allowing the \textit{OrderManagement} microservice to communicate with the \textit{InventoryManagement} microservice via remote synchronous service calls using HTTP requests.

\subsubsection*{Asynchronous solution} The asynchronous solution using \textit{Apache Kafka} is shown in Listings~\ref{final_service_call_async_doc} and~\ref{final_service_call_async_doc2}.
\begin{lstlisting}[language=Java, label=final_service_call_async_doc, caption={[Changing local call to an asynchronous service call in the \textit{OrderManagement} microservice]Asynchronous solution for the \emph{OrderManagement} microservice implementation.}]
// Candidate for the OrderManagement microservice
public class OrderEvent {
    private Order order;
    // Constructors, getters, and setters
}
public class OrderUpdatedEvent {
    private Order order;
    // Constructors, getters, and setters
}
@Component
public class KafkaOrderEventProducer {
    private final KafkaTemplate<String, OrderEvent> kafkaTemplate;
    private final String topic;

    public KafkaOrderEventProducer(KafkaTemplate<String, OrderEvent> kafkaTemplate, @Value("${kafka.topic}") String topic) {
        this.kafkaTemplate = kafkaTemplate;
        this.topic = topic;
    }

    public void publishOrderEvent(OrderEvent orderEvent) {
        kafkaTemplate.send(topic, orderEvent);
    }
}
public class OrderProcessor {
    private final KafkaOrderEventProducer eventProducer;

    public OrderProcessor(KafkaOrderEventProducer eventProducer) {
        this.eventProducer = eventProducer;
    }

    public void processOrder(Order order) {
        // Business logic for processing the order
        // ...

        // Publish event asynchronously
        OrderEvent event = new OrderEvent(order);
        eventProducer.publishOrderEvent(event);
    }
}
\end{lstlisting}
\begin{lstlisting}[language=Java, label=final_service_call_async_doc2, caption={[Changing local call to an asynchronous service call in the \textit{OrderManagement} microservice]Asynchronous solution for the\emph{InventoryManagement} microservice implementation.}]
// Candidate for the InventoryManagement microservice
@Service
public class InventoryService {

    public void updateInventory(Order order) {
        // Inventory update logic
        // ...
    }
}

@Component
public class OrderEventListener {

    private final InventoryService inventoryService;

    public OrderEventListener(InventoryService inventoryService) {
        this.inventoryService = inventoryService;
    }

    @KafkaListener(topics = "${kafka.topic}")
    public void handleOrderEvent(OrderEvent event) {
        inventoryService.updateInventory(event.getOrder());
    }
}
\end{lstlisting}

We first introduced a new class, \textit{OrderEvent}, to encapsulate the data associated with an order. Next, we implemented a Kafka producer component, \textit{KafkaOrderEventproducer}, responsible for publishing instances of \textit{OrderEvent} to a designated Kafka topic. On the consumer side, the \textit{OrderEventListener} subscribes to this topic and reacts to incoming events by invoking the \textit{updateInventory} method of the \textit{InventoryService}, thereby performing the necessary inventory updates.

The \textit{OrderProcessor} class was modified to use the \textit{KafkaOrderEventproducer} instead of making a direct method call to the inventory component. This change decouples the order and inventory logic, enabling asynchronous communication between the two services. As a result, the system transitions from tightly coupled direct method calls to an event-driven architecture.

\subsection{Move Foreign-key Relationship to Code} \label{moveforeigndoc}
\subsubsection*{Context and Motivation}
When two entities are related and dependent on one another, their relationship is often represented in a relational database using foreign-key constraints. These relationships - such as One-to-One, Many-to-One, or One-to-Many - enable referential integrity and simplify querying through joins.

However, when decomposing a monolithic system into microservices, each service must own and manage its own data. This means that when we are planning to extract a service and realise that such domain entities should be in different microservices, we also realise that the database tables representing these entities must be separated into distinct schemas, each controlled by its respective service. If a foreign-key constraint exists between these tables, it becomes a barrier to service extraction.

In particular, when we want one service to own the table containing the foreign key and another to own the referenced table, before we can move the tables to different databases, we must eliminate this explicit foreign-key constraint, while still being able to relate the data in the two services to ensure the same functionality. Therefore, in addition to eliminating any foreign-key constraint from the database, any database query that previously joined the two tables will have to be reimplemented by service code using explicit inter-service communication.

This refactoring is essential to eliminate tight data-level coupling and enable independent evolution of services.

\subsubsection*{Example}
Consider the above-mentioned system that manages order processing and inventory tracking, which was initially implemented as a monolith. 

A scenario illustrating the need for this refactoring is as follows: \textit{the \emph{Order} entity has a Many-to-One relationship with the \emph{Customer} entity, implemented via a foreign-key constraint on the \emph{customer\_id} column in the orders table}.

Because this relationship is enforced at the database level, it assumes both tables exist within the same schema and runtime environment. This dependency means:
\begin{itemize}
    \item Extracting \emph{OrderManagement} into a separate microservice would break the foreign-key constraint.
    \item The relationship between orders and customers is hidden within database joins, rather than exposed through service interfaces.
    \item The \emph{Customer} table cannot be moved to a separate database without redesigning how \emph{OrderManagement} accesses customer data.
\end{itemize}
To enable microservice extraction, the foreign-key constraint is removed, and the \emph{OrderManagement} service uses the \emph{customer\_id} as a reference. When customer data is needed, it is retrieved via a service call to \emph{CustomerManagement}. This transformation makes the dependency explicit and allows each service to evolve independently.

\subsubsection*{Strategy}
To decouple the data models and prepare for service extraction, we remove the foreign-key constraint from the database and shift the responsibility for maintaining the relationship to the application layer.

This involves:
\begin{itemize}
    \item Eliminating the foreign-key constraint from the schema.
    \item Replacing database joins with service calls that retrieve related data from the owning service.
    \item Using identifiers (e.g., customer\_id) as references, without enforcing referential integrity at the database level.
    \item Optionally introducing caching or denormalization strategies to reduce the overhead of repeated service calls.
\end{itemize}

This way, we preserve the logical relationship between the entities while allowing each microservice to manage its own data independently.

\subsubsection*{Benefits} 
This refactoring supports:
\begin{itemize}
    \item Greater autonomy in local transactions: by removing the cross-table constraints, each service can manage its own transactions without needing distributed coordination.
    \item Reduced data-level coupling between services, as we are decoupling the physical data models, each service can evolve its schema independently.
    \item Improved local testability and maintainability: with no reliance on external table joins, services can be tested in isolation.
    \item Support for heterogeneous technologies, with this decoupling, services can adopt different databases or storage engines without compatibility concerns.
\end{itemize}
\subsubsection*{Challenges}
This approach introduces some challenges, namely:
\begin{itemize}
    \item The loss of referential integrity guarantees.
    \item Ensuring data consistency across microservices.
    \item Increased complexity of distributed transactions as they now require coordination across services.
\end{itemize}

\subsubsection*{Mechanics}
The following steps can either be performed after breaking the code dependency or with the code breakage in mind. When referring to services, we describe the intended future service boundaries — these services do not need to be fully implemented yet.
\begin{enumerate}
    \item Remove the foreign-key constraint from the table that has it.
    \item If you have classes in your code representing the domain entities that translate into the database tables that used to be subject to the foreign-key constraint, create an attribute in one of those classes to represent the other entity involved in the relationship. This new attribute should be translated into the column in this entity's table that used to have the foreign-key constraint. When retrieving data, we will no longer use it to join tables, but as a query filter.
    \item Separate the tables according to ownership boundaries. Assign each table to the domain that conceptually owns it (typically aligned with the future microservice boundaries). This separation doesn’t need to be physical at this stage; it can be represented through logical partitioning, such as distinct database views, schemas, or access layers. The goal is to make ownership explicit and prepare for eventual physical separation if needed.
    \item Create a data access interface for each of these databases that implements the methods of data manipulation.
    \item Identify the methods that previously accessed or manipulated data across both tables and refactor them to use the newly created interfaces, ensuring that each method interacts only with the data owned by its domain.
    \item When you separate the services, don’t forget to use the previous refactoring to \textbf{``Replace Method Call with Service Call''(~\ref{changeCalldoc})} to change these local methods to service calls, passing the primary key as a parameter to retrieve related data.
\end{enumerate}

\textbf{Notes:}
\begin{itemize}
    \item We may need to remove code annotations when using specific programming languages, frameworks or ORMs that use them.
    \item The way we join the information of these two entities is no longer through a join query, so data consistency has to be a concern. Do not forget to implement mechanisms to guarantee data integrity and consistency~\footnote{\scriptsize{Complementary explanation available in the related documentation (check item C.7.1): \url{https://github.com/RitaPeixoto/Migration-of-Monoliths-to-Microservices-Survey_replication_package/blob/main/catalogue_of_refactorings.pdf}}}.
    \item Be aware that the latency of requests increases as we transform the database calls into service calls. Design your interactions carefully to minimize performance impact, possibly using caching or batching where appropriate.
\end{itemize}

\subsubsection*{Example of application} In the Listings ~\ref{initial_moveforeign_doc} and ~\ref{initial_moveforeign_doc2}, we can see the current code of the monolith where the 
\textit{Order} entity has a \textit{ManyToOne} relationship with the entity \textit{Customer}.
This relationship is implemented using a foreign-key constraint on the \textit{customer\_id} column in the orders table.

\begin{lstlisting}[language=Java, label=initial_moveforeign_doc, caption={The \textit{Order} related classes.}]
// Candidate for the OrderManagement microservice
@Entity
@Table(name = "orders")
public class Order {

    @Id
    @GeneratedValue(strategy = GenerationType.IDENTITY)
    private Long id;

    @ManyToOne
    @JoinColumn(name = "customer_id")
    private Customer customer;

    //  ...
}
@Service
public class OrderService {

    private final OrderRepository orderRepository;

    public OrderService(OrderRepository orderRepository) {
        this.orderRepository = orderRepository;
    }

    public void processOrder(Order order) {
        // Perform business logic
        Customer customer = order.getCustomer();

        // Use customer data for processing (e.g., validate, enrich, etc.)

        orderRepository.save(order);
    }
}

@Repository
public interface OrderRepository extends JpaRepository<Order, Long> {
    // Order-related methods for data manipulation
}
\end{lstlisting}

\begin{lstlisting}[language=Java, label=initial_moveforeign_doc2, caption={The \textit{Customer} class.}]
// Candidate for the CustomerManagement microservice
@Entity
@Table(name = "customers")
public class Customer {

    @Id
    @GeneratedValue(strategy = GenerationType.IDENTITY)
    private Long id;

    // Other properties, constructors, getters, and setters
}
\end{lstlisting}
The \textit{OrderService} class depends on the \textit{OrderRepository} interface for data access and manipulation. The \textit{processOrder} method accesses the \textit{Customer} entity directly via the \textit{getCustomer()} method on the \textit{Order} object.

As these entities will belong to different microservices in the future, we need to refactor this dependency and move the foreign key relationship into the code.

Listing~\ref{final_moveforeign_doc} shows the code after applying the refactoring.
\begin{lstlisting}[language=Java, label=final_moveforeign_doc, caption={[Move Foreign-key constraint dependency between \textit{Order} and \textit{Customer} to code]Updated Order Entity and Service Layer After Refactoring.}]
// Candidate for the OrderManagement microservice
@Entity
@Table(name = "orders")
public class Order {

    @Id
    @GeneratedValue(strategy = GenerationType.IDENTITY)
    private Long id;

    private Long customerId;

    // Other properties, constructors, getters, and setters
}

@Service
public class OrderService {

    private final OrderRepository orderRepository;
    private final CustomerRepository customerRepository;

    public OrderService(OrderRepository orderRepository, CustomerRepository customerRepository) {
        this.orderRepository = orderRepository;
        this.customerRepository = customerRepository;
    }

    public void processOrder(Order order) {
        // Retrieve customer data using the customerId
        Long customerId = order.getCustomerId();
        Customer customer = customerRepository.findById(customerId)
            .orElseThrow(() -> new IllegalArgumentException("Customer not found"));

        // Use customer data for business logic

        orderRepository.save(order);
    }
}
[...]
\end{lstlisting}
We removed the foreign-key constraint from the \textit{Order} table referencing the \textit{Customer} table. The \textit{Order} entity now contains a simple \textit{customerId} field to represent the association. This field is no longer used for database joins but serves as a reference for service-level lookups.

Each table is now assigned to a separate database: the \textit{orders\_db} for the entity \textit{Order} and the \textit{customers\_db} for the entity \textit{Customer}.

We created the interfaces for data manipulation. The \textit{OrderService} now depends on both \textit{OrderRepository} and \textit{CustomerRepository}, each targeting a distinct database. The \textit{processOrder} method retrieves the \textit{Customer} entity using the \textit{customerId} stored in the \textit{Order} object. This lookup replaces the implicit ORM join and makes the relationship explicit in the service layer.

\subsection{Replicate Data Across Microservices}\label{replicate_data_doc}
\subsubsection*{Context and Motivation}
In a microservices architecture, each service should own and manage its own data. However, in practice, different services often need access to the same data.  If multiple services query or manipulate the same shared database, they will not be entirely independent, as one microservice will also manage the data of another.

To preserve service independence while still allowing access to shared data, we replicate the necessary data across services.  One of the services is the data owner and source of truth, while others maintain read-only copies of the data they need to access and manipulate. This replication can be implemented using various strategies, such as database-level replication, event sourcing, or change data capture.

Ideally, replication should avoid distributed transactions and embrace eventual consistency, although synchronous replication may be used in specific scenarios. This approach allows services to operate independently while maintaining access to relevant data.

\subsubsection*{Example}
Consider the above-mentioned system that manages order processing
and inventory tracking, which was initially implemented as a monolith.

A scenario illustrating the need for this refactoring is as follows: \textit{the \textit{OrderProcessor} class, belonging to the \textit{OrderManagement} domain, needs to validate product availability before confirming an order. To do this, it directly queries the inventory data managed by the \textit{InventoryService} class under the \textit{InventoryManagement} domain}.

Because this interaction is implemented through direct access to shared database tables or internal method calls, it assumes both components operate within the same runtime and data store. This dependency introduces several limitations:
\begin{itemize}
    \item Extracting \textit{OrderProcessor} into a separate microservice would break its ability to access inventory data directly.
    \item The relationship between order logic and inventory state is hidden within internal queries, rather than exposed through well-defined interfaces or protocols.
    \item \textit{InventoryService} cannot be moved to a separate service without redesigning how \textit{OrderProcessor} obtains inventory information.
\end{itemize}
To enable microservice extraction and preserve functionality, the inventory data needed by \textit{OrderManagement} is replicated from \textit{InventoryManagement}. This is achieved by publishing domain events such as \textit{StockLevelUpdated} or \textit{ProductOutOfStock}, which \textit{OrderManagement} subscribes to and uses to maintain a local copy of relevant inventory data.

This transformation allows \textit{OrderProcessor} to make decisions based on locally stored inventory snapshots, without querying \textit{InventoryService} directly. It decouples the services, supports independent deployment, and embraces eventual consistency, shifting from shared data access to replicated, service-owned data.

\subsubsection*{Strategy}
Replicating data across microservices is a strategic move to balance autonomy with operational coherence. The first step is to establish clear data ownership: one service must be recognized as the authoritative source, responsible for maintaining a given dataset. Other services that rely on this data do not query it directly, but instead maintain their own local copies tailored to their needs. There are several ways to achieve this replication, each suited to different architectural contexts, but regardless of the method chosen, the replication strategy should be designed with resilience in mind. Services must tolerate latency, operate with eventual consistency, and remain functional even if the source service is temporarily unavailable. This ensures that data dependencies do not become bottlenecks, and that each microservice can evolve and scale independently.

\subsubsection*{Benefits} 

Replicating data across services offers:
\begin{itemize}
    \item Reduced coupling: as we are replacing direct queries with replicated data access, we are explicitly decoupling services at the data level.
    \item Improved scalability: replication allows services to optimize reads locally.
    \item Improved resilience: services can continue operating with local data even if the source service is down.
\end{itemize}

\subsubsection*{Challenges} 

However, it can introduce challenges such as:
\begin{itemize}
    \item Data synchronization: it is hard to keep replicated data in sync with the source.
    \item Update conflicts: the replicated data must be treated as read-only to avoid conflicts.
    \item Replication strategy complexity.
    \item Storage overhead: we are maintaining multiple copies of data.
\end{itemize}

\subsubsection*{Mechanics}
\begin{enumerate}[itemsep=1ex]
    \item Determine which service will be the owner of the shared data, and, therefore, the source of truth. 
    \item Choose a replication strategy: \begin{enumerate}
        \item Using database-level replication channels: if supported by the engine, you can create one or more replication channels between it and the shared data source.
        \item Using event sourcing to publish domain events that other services can consume. Event sourcing is a method of storing (or communicating) data which facilitates data replication because events may be easily repeated. It is a way to keep eventual consistency. It holds events that are frequently objects, and because event sourcing does not need to know its consumers, other technologies can be utilized concurrently (for more on event sourcing, check Martin Fowler's article~\cite{fowler_eventSourcing}).
        \item Using \textbf{``Change Data Capture'' refactoring~\footnote{\scriptsize{Change data capture is a technique to identify and record the changes that occur in a database. It delivers these changes in real-time to different target systems, enabling the synchronization of data to the services that need it when a database change occurs~\cite{datacapture_ibm}. Complementary explanation available in the related documentation (check item C.5.12): \url{https://github.com/RitaPeixoto/Migration-of-Monoliths-to-Microservices-Survey_replication_package/blob/main/catalogue_of_refactorings.pdf}}}} to propagate updates from the source database.
    \end{enumerate}
    \item In the consuming service, define entities or views that represent the replicated data.
    \item Implement mechanisms to receive and process updates from the source service (e.g., event listeners, CDC consumers).
    \item Store the replicated data in the consumer’s database, ensuring it is treated as read-only.
    \item Modify methods that previously queried the source database to use the local replicated data instead.
\end{enumerate}
\textbf{Important:} Implement safeguards to handle out-of-sync data, retries, and reconciliation if needed.
\subsubsection*{Example of application}
This example demonstrates how to utilise Event Sourcing to Replicate Data Across Microservices.

Listing~\ref{replicate_initial} shows the current code of the monolith, where the \textit{OrderService} class directly queries inventory data to validate product availability before confirming an order. 

\begin{lstlisting} [language=Java, label=replicate_initial, caption={\textit{OrderService} directly queries inventory data, creating a runtime and data dependency on \textit{InventoryService}.}]
// Candidate for the OrderManagement microservice
@Entity
@Table(name = "orders")
public class Order {

    @Id
    @GeneratedValue(strategy = GenerationType.IDENTITY)
    private Long id;

    @ManyToOne
    @JoinColumn(name = "user_id")
    private User user;

    // Other properties, constructors, getters, and setters
    // ...
}

@Service
public class OrderService {

    private final OrderRepository orderRepository;
    private final InventoryRepository inventoryRepository;

    public OrderService(OrderRepository orderRepository, InventoryRepository inventoryRepository) {
        this.orderRepository = orderRepository;
        this.inventoryRepository = inventoryRepository;
    }

    public void processOrder(Order order) {
        InventoryItem item = inventoryRepository.findByProductId(order.getProductId());

        if (item.getStockLevel() > 0) {
            orderRepository.save(order);
        } else {
            throw new OutOfStockException();
        }
    }
}
\end{lstlisting}
To resolve this, \textit{InventoryManagement} becomes the owner of inventory data and publishes domain events such as \textit{StockLevelUpdatedEvent} whenever stock levels change. 

Listing~\ref{replicate_inventory_doc} shows the code of the \textit{InventoryManagement} microservice after implementing the event sourcing.
\newpage

\begin{lstlisting} [language=Java, label=replicate_inventory_doc, caption={\textit{InventoryService} publishes a \textit{StockLevelUpdatedEvent} whenever stock levels change.}]
// Candidate for the InventoryManagement microservice
@Service
public class InventoryService {

    private final InventoryRepository inventoryRepository;
    private final EventPublisher eventPublisher;

    public InventoryService(InventoryRepository inventoryRepository, EventPublisher eventPublisher) {
        this.inventoryRepository = inventoryRepository;
        this.eventPublisher = eventPublisher;
    }

    public void updateStock(String productId, int newStockLevel) {
        InventoryItem item = inventoryRepository.findByProductId(productId);
        item.setStockLevel(newStockLevel);
        inventoryRepository.save(item);

        StockLevelUpdatedEvent event = new StockLevelUpdatedEvent(productId, newStockLevel);
        eventPublisher.publish(event);
    }
}
\end{lstlisting}

\textit{OrderManagement} subscribes to these events and maintains a local, read-only copy of the inventory data it needs. Listing~\ref{replicate_order_doc} shows the code of the \textit{OrderManagement} service that subscribes to this event and, when an event is received, updates the local record with the replicated inventory data.

\begin{lstlisting} [language=Java, label=replicate_order_doc, caption={\textit{OrderManagement} maintains a local read-only copy of inventory data and uses it to validate orders.}]
// Candidate for the OrderManagement microservice
@Service
public class InventoryReplicationService {

    private final InventorySnapshotRepository snapshotRepository;

    public InventoryReplicationService(EventSubscriber eventSubscriber, InventorySnapshotRepository snapshotRepository) {
        this.snapshotRepository = snapshotRepository;
        eventSubscriber.subscribe(StockLevelUpdatedEvent.class, this::handleStockUpdate);
    }

    private void handleStockUpdate(StockLevelUpdatedEvent event) {
        InventorySnapshot snapshot = new InventorySnapshot(event.getProductId(), event.getStockLevel());
        snapshotRepository.save(snapshot);
    }
}

@Service
public class OrderService {

    private final OrderRepository orderRepository;
    private final InventorySnapshotRepository snapshotRepository;

    public OrderService(OrderRepository orderRepository, InventorySnapshotRepository snapshotRepository) {
        this.orderRepository = orderRepository;
        this.snapshotRepository = snapshotRepository;
    }

    public void processOrder(Order order) {
        InventorySnapshot snapshot = snapshotRepository.findByProductId(order.getProductId());

        if (snapshot != null && snapshot.getStockLevel() > 0) {
            orderRepository.save(order);
        } else {
            throw new OutOfStockException();
        }
    }
}
\end{lstlisting}

This replication allows \textit{OrderService} to make decisions based on its own local snapshot of inventory, without querying another service or database directly. It preserves service autonomy, supports independent scaling, and embraces eventual consistency.
\subsection{Split Database Across Microservices}\label{splitdatabase_doc}
\subsubsection*{Context and Motivation}
When extracting services from a monolithic system, it is common to encounter database tables that aggregate data belonging to multiple business domains. These tables are often accessed and manipulated by different components that will eventually become independent microservices.

This shared access creates a significant obstacle to service decomposition. Splitting a monolithic database is not trivial, it requires careful analysis of ownership, access patterns, and dependencies.

\subsubsection*{Example}
Consider the above-mentioned system that manages order processing and inventory tracking, which was initially implemented as a monolith.

A scenario illustrating the need for this refactoring is as follows: \textit{the Product table contains both inventory-related fields (e.g., stockQuantity, warehouseLocation) and pricing-related fields (e.g., price, discount). In the monolithic system, both the \textit{OrderManagement} and \textit{InventoryManagement} components access and update this table directly.}

After decomposition, both \textit{OrderManagement} and \textit{InventoryManagement} are extracted into separate microservices. However, they continue to rely on the same \textit{Product} table and update overlapping columns, such as price, which is used by \textit{OrderManagement} to calculate totals, and by \textit{InventoryManagement} to adjust pricing based on stock levels.
This setup introduces several limitations:
\begin{itemize}
    \item There is no clear ownership of the \textit{price} field, making schema evolution and business logic changes risky.
    \item Concurrent updates from both services can lead to conflicts and inconsistencies.
    \item Extracting either service without redesigning access to the shared table would break functionality and violate service autonomy.
\end{itemize}
To enable microservice extraction and preserve functionality, we must assign ownership of the shared columns to one microservice. In this case, \textit{InventoryManagement} becomes the owner of the \textit{Product} table and its pricing logic. \textit{OrderManagement}, which still needs to update pricing in specific scenarios (e.g., applying discounts), must now do so via a service call to \textit{InventoryManagement}.

This approach ensures that only one service owns and updates the data, while others interact through well-defined interfaces, shifting from shared table updates to coordinated service-mediated access.

\subsubsection*{Strategy}
The goal is to isolate data ownership so that each microservice manages only the data it is responsible for, which involves:
\begin{itemize}
    \item Identifying which tables are exclusively used by a single microservice and moving them directly.
    \item Analyzing shared tables to determine column ownership and access patterns.
    \item Applying different strategies depending on whether services read or write to the same columns.
\end{itemize}

There are three common scenarios for this refactoring:
\begin{itemize}
    \item Shared table, distinct columns.
    \item Shared table, shared columns.
    \item Shared table, one service writes, and one only reads.
\end{itemize}

Depending on the scenario, a specific strategy shall be applied.

\subsubsection*{Benefits}
Splitting databases across microservices can bring some benefits, such as:
\begin{itemize}
    \item Clear data ownership and schema boundaries.
    \item Improved scalability and database flexibility.
\end{itemize}

\subsubsection*{Challenges}
However, this refactoring also presents considerable challenges, including:
\begin{itemize}
    \item Referential integrity in a distributed environment: foreign-key constraints don’t work across service boundaries, so you need to enforce relationships in code or through service calls.
    \item Increased operational complexity: migrating data, adapting queries, and ensuring consistency during the transition can be technically demanding.
    \item When multiple services update the same columns, deciding ownership and coordination becomes complex.
\end{itemize}

\subsubsection*{Mechanics}
\begin{enumerate}[itemsep=1ex]
    \item Identify the tables used exclusively by each microservice and move them directly to that microservice's database.
    \item Analyze shared tables to determine column ownership and access pattern.
    \item Choose the strategy to apply based on services access patterns:
    \begin{enumerate}
        \item If two microservices access the same database table but manipulate different columns:
        \begin{enumerate}
            \item Option 1: Replicate the table accross both microservices using \textbf{``Replicate Data Across Microservices''} (Section~\ref{replicate_data_doc}) and use a data replication mechanism to keep it consistent.
            \item Option 2: Split the table into two separate tables, each containing only the columns relevant to its respective service. 
            \item In each component, include the corresponding table and adapt the code to use its own table.
            \item If foreign key relationships existed in the monolith, replace them with service-level references using \textbf{``Move Foreign-key Relationship to Code''} refactoring (Section~\ref{moveforeigndoc}).
        \end{enumerate}
        \item If two microservices access the same database table and update the same columns:
        \begin{enumerate}
            \item Option 1: Replicate the data for both microservices using \textbf{``Replicate Data Across Microservices''} (Section~\ref{replicate_data_doc}) and use a data replication mechanism to keep it consistent
            \item Option 2: Assign ownership of the shared columns to one microservice.
            \item Make the other microservice interact with the owning service via a service call to update this column. 
            \item To migrate incrementally, first refactor the monolith so that the non-owning component updates the data via a method call. Later, replace this with a remote service call using the refactoring \textbf{``Replace Method Call with Service Call''} (Section~\ref{changeCalldoc}).
        \end{enumerate}
        \item If one microservice has read-write access to a table and another only reads from it:
        \begin{enumerate}
            \item Assign ownership of the table to the read-write microservice.
            \item The read-only microservice should retrieve the necessary data via a service call to the owning microservice.
            \item Use the refactoring \textbf{``Replace Method Call with Service Call''}  (Section~\ref{changeCalldoc}) to replace direct data access with a well-defined interface. 
        \end{enumerate}
    \end{enumerate}
\end{enumerate}
\textbf{Note:} Guarantee data consistency~\footnote{\scriptsize{Complementary explanation available in the related documentation (check item C.7.1): \url{https://github.com/RitaPeixoto/Migration-of-Monoliths-to-Microservices-Survey_replication_package/blob/main/catalogue_of_refactorings.pdf}}}

\subsubsection*{Example of application}
In this example of application we illustrate scenario (b) without data replication. 

Listing~\ref{orderservice} shows a part of the code of the monolith, where the \textit{OrderService} class (and, not shown, the \textit{InventoryService} class) interact with the \textit{Product} table directly. \textit{OrderService} applies discounts during promotions, while \textit{InventoryService} will adjust prices based on stock levels or supplier changes.

\begin{lstlisting}[language=java, label=orderservice, caption=\textit{OrderService} directly updates the \textit{Product} table - creating shared write access with \textit{InventoryService}.]
// Candidate for the OrderManagement microservice
@Service
public class OrderService {

    private final ProductRepository productRepository;

    public OrderService(ProductRepository productRepository) {
        this.productRepository = productRepository;
    }

    public void applyDiscount(Long productId, BigDecimal discount) {
        Product product = productRepository.findById(productId);
        product.setDiscount(discount);
        productRepository.save(product);
    }
}
\end{lstlisting}

To decouple the services and clarify ownership, we assign the \textit{Product} table to the \textit{InventoryManagement} microservice, making it the owner of pricing-related data. The \textit{OrderManagement} microservice, which still needs to update pricing in specific scenarios, now does so via a remote service call to \textit{InventoryManagement}. We remove the direct database updates from \textit{OrderManagement} microservice to the shared columns in the \textit{Product} table and change them to make service calls to the API provided by \textit{InventoryManagement} microservice whenever updates to the shared columns related to inventory management, are required.

Listing~\ref{split_inventory_doc} shows the code on the \textit{InventoryManagement} microservice side and Listing~\ref{split_order_doc} shows the code on the \textit{OrderManagement} microservice side.

\begin{lstlisting}[language=java, label=split_inventory_doc, caption=\textit{InventoryManagement} exposes an HTTP endpoint to update product discounts centralizing ownership of pricing data.]
// Candidate for the InventoryManagement microservice
@RestController
@RequestMapping("/api/products")
public class InventoryController {

    private final ProductRepository productRepository;

    public InventoryController(ProductRepository productRepository) {
        this.productRepository = productRepository;
    }

    @PutMapping("/{productId}/discount")
    public ResponseEntity<Void> updateDiscount(@PathVariable Long productId,
                                               @RequestBody BigDecimal discount) {
        Product product = productRepository.findById(productId);
        product.setDiscount(discount);
        productRepository.save(product);
        return ResponseEntity.ok().build();
    }
}
\end{lstlisting}

\begin{lstlisting}[language=java, label=split_order_doc, caption=\textit{OrderService} delegates discount updates to InventoryManagement via a service call -  removing direct access to the shared table.]
// Candidate for the OrderManagement microservice
@Service
public class InventoryClient {

    private final RestTemplate restTemplate;

    public InventoryClient(RestTemplate restTemplate) {
        this.restTemplate = restTemplate;
    }

    public void updateDiscount(Long productId, BigDecimal discount) {
        String url = "http://inventory-service/api/products/" + productId + "/discount";
        restTemplate.put(url, discount);
    }
}

@Service
public class OrderService {

    private final InventoryClient inventoryClient;

    public OrderService(InventoryClient inventoryClient) {
        this.inventoryClient = inventoryClient;
    }

    public void applyDiscount(Long productId, BigDecimal discount) {
        inventoryClient.updateDiscount(productId, discount);
    }
}
\end{lstlisting}

 With this refactoring, the \textit{InventoryManagement} microservice has ownership over the inventory-related data, while the \textit{OrderManagement} microservice interacts with the \textit{InventoryManagement} microservice through service calls to update the shared inventory columns. This way, each microservice focuses on its specific responsibilities.

\subsection{Create Data Transfer Object}\label{dto_doc}
\subsubsection*{Context and Motivation}
This refactoring is commonly necessary when we extract a service and there is a relationship between entities that will belong to different microservices. Components often need to interact with each other to perform their operations and these interactions frequently involve multiple pieces of related data, such as customer details, product information, or configuration parameters, that are directly accessible.

However, when transitioning to a microservices architecture, these entities are split across service boundaries. Each microservice owns and manages its own data, and direct access to related entities in other services is no longer possible. Despite this, services still need to exchange structured data to perform operations collaboratively.

\subsubsection*{Example}
Consider the above-mentioned system that manages order processing and inventory tracking, which was initially implemented as a monolith.

A scenario illustrating the need for this refactoring is as follows: \textit{the OrderService class, belonging to the OrderManagement domain, exposes a method called getOrderDetails, which returns an Order entity. This entity contains nested references to other domain objects, such as Customer and Product, and is used directly by other components or services.}

Because this interaction is implemented by returning a full domain object, it assumes that consumers of the service operate within the same runtime and share the same domain model. This dependency introduces several limitations:
\begin{itemize}
    \item Extracting \textit{OrderService} into a separate microservice would break its ability to share data without exposing internal domain logic.
    \item The relationship between order logic and customer/product data is tightly coupled to the structure of the \textit{Order} entity.
    \item Consumers of the service must understand and depend on the internal model of \textit{Order}, making independent evolution difficult.
\end{itemize}

To enable microservice extraction and preserve autonomy, the data returned by \textit{OrderService} must be encapsulated in a Data Transfer Object (DTO). This DTO contains only the necessary fields for communication and is decoupled from the internal domain model.

This transformation allows \textit{OrderService} to expose a stable, serializable structure for external consumers, while retaining the flexibility to evolve its internal model independently, capturing the essence of the refactoring need: shifting from domain model exposure to structured data transfer.

\subsubsection*{Strategy} The goal is to decouple internal domain models from external communication formats. Therefore, we shall create a Data Transfer Object (DTO) that aggregates all the necessary data into a single, serializable structure. This object is designed specifically for communication between services and is decoupled from the internal domain models of either side.

A DTO is designed specifically for data exchange between services and should:
\begin{itemize}
    \item Contain only the fields required for the operation in question.
    \item Be serializable for transmission over the network (e.g., via HTTP, messaging, or RPC).
    \item Be maintained independently of domain models to preserve service autonomy and avoid tight coupling.
\end{itemize}

DTOs are especially useful when:
\begin{itemize}
    \item A service needs to expose a simplified or enriched view of its internal data.
    \item Multiple pieces of related data must be bundled into a single response.
    \item The consuming service should not depend on the internal structure of the source domain.
\end{itemize}

\subsubsection*{Benefits} The main benefits of this refactoring include:
\begin{itemize}
    \item Bundles related data into a single structure, reducing the number of calls between microservices, which decreases latency.
    \item Reduced coupling: Services no longer depend on each other's domain models.
    \item Enhances flexibility by decoupling the DTO from domain models.
\end{itemize}

\subsubsection*{Challenges} However, its challenges include:
\begin{itemize}
    \item Maintaining data consistency. It is challenging to ensure that the DTO accurately reflects up-to-date data from the source service.
    \item Managing the complexity of defining, transforming, and maintaining data transfer objects, without impacting performance.
    \item Managing DTO changes over time without breaking consumers.
\end{itemize}

\subsubsection*{Mechanics}
\begin{enumerate}
    \item Identify the data to be transferred: determine which fields are needed by the consuming service. Avoid exposing internal domain logic or unnecessary attributes.
    \item Define the DTO class: create a new class (Data Transfer Object - DTO) that contains only the required fields for the communication between the services. Ensure it is serializable and independent of domain entities.
    \item Transform domain entities into DTOs: in the service layer, convert domain objects into DTOs before returning or transmitting them.
    \item Update service interfaces: replace method signatures that return domain entities with versions that return DTOs.
    \item Maintain DTO evolution independently: as requirements change, evolve the DTO without affecting the internal domain model. This preserves flexibility and autonomy.
\end{enumerate}
\subsubsection*{Example of application} 
Originally, the \textit{OrderService} class in the \textit{OrderManagement} microservice exposes a method called \textit{getOrderDetails}, which returns an \textit{Order} entity. This entity includes nested references to other domain objects such as \textit{Customer} and \textit{Product}, and is used directly by external consumers. An example of this implementation can be seen in Listing~\ref{order_doc}.

However, once the system is decomposed into microservices, returning a full domain entity becomes problematic. The \textit{Order} class may contain internal logic or relationships that are irrelevant, or even inaccessible, to other services. Moreover, sharing domain models across service boundaries introduces tight coupling and hinders independent evolution.

In the \textit{getOrderDetails} method from \textit{Order} microservice class, an object of type \textit{Order} is being sent through the communication. However, we want to create a Data Transfer Object that can hold the necessary data in a call to this method that contains more than information only present in the \textit{Order} class. This way, the services will not have to share the same entity because we are encapsulating the specific data for communication, creating an abstraction.
 
\begin{lstlisting} [language=Java, label=order_doc, caption=The \textit{OrderService} returns a full \textit{Order} entity - exposing internal relationships and structure.]
// Candidate for the OrderManagement microservice
@Entity
public class Order {
    @Id
    @GeneratedValue(strategy = GenerationType.IDENTITY)
    private Long id;
    
    private String customerName;

    @OneToMany
    private List<Product> products;
    
    // Other fields and relationships
    
    // Constructors, getters, and setters
}

@Service
public class OrderService {
    private final OrderRepository orderRepository;

    public OrderService(OrderRepository orderRepository) {
        this.orderRepository = orderRepository;
    }

    public Order getOrderDetails(Long orderId) {
        return orderRepository.findById(orderId);
    }
}
\end{lstlisting}
To resolve this, we introduce a Data Transfer Object (DTO) named \textit{OrderDTO}, which encapsulates only the necessary data for external communication. This DTO abstracts the internal structure of the \textit{Order} entity and provides a stable format for transferring order-related information. The transformation is shown in 
Listing~\ref{dto_ex_doc}.

\begin{lstlisting} [language=Java, label=dto_ex_doc, caption=The \textit{OrderService} now returns an \textit{OrderDTO} - decoupled from the internal domain model and tailored for communication.]
// Candidate for the OrderManagement microservice
public class OrderDTO {
    private Long orderId;
    private String customerName;
    private List<String> products;
    // Other fields as needed

    // Constructors, getters, and setters
}

@Service
public class OrderService {
    private final OrderRepository orderRepository;

    public OrderService(OrderRepository orderRepository) {
        this.orderRepository = orderRepository;
    }

    public OrderDTO getOrderDetails(Long orderId) {
        Order order = orderRepository.findById(orderId);
       
        OrderDTO orderDTO = new OrderDTO();
        orderDTO.setOrderId(order.getId());
        orderDTO.setCustomerName(order.getCustomer().getName());
        orderDTO.setProducts(order.getProducts().stream().map(Product::getName)
            .collect(Collectors.toList()));
        // Set other fields as needed

        return orderDTO;
    }
}
\end{lstlisting}

We define a new class representing the DTO and declared the necessary fields to hold the data. In the future, more fields can be added to this DTO as they correspond to the transferred data. Then, we transform the data being transferred into the DTO. 

The \textit{getOrderDetails} method is then updated to return an instance of \textit{Order} instead of the original \textit{Order} entity. This transformation ensures that the consuming services receive only the relevant data, without depending on the internal domain model.

The DTO can evolve independently from the domain entity, allowing new fields to be added as communication needs change. It provides a standard format for transferring the data of orders between services.

\subsection{Break Data Type Dependency}

\subsubsection*{Context and Motivation}
In monolithic systems, it is common for components to share data types across different business domains. This dependency can appear in attributes types, parameter types, return types, and even method attributes types. These shared types often reflect implicit coupling between functionalities that, while logically distinct, are tightly bound through code-level dependencies.

When transitioning to a microservice architecture organized by business capabilities, such data type dependencies can become problematic. A microservice may require access to a type defined in another domain, even if only for a small part of its operation.

\subsubsection*{Example} Consider the above-mentioned system that manages order processing
and inventory tracking, which was initially implemented as a monolith. 

A scenario illustrating the need for this refactoring is as follows: \textit{both the \textit{OrderManagement} and \textit{InventoryManagement} components rely on a shared \textit{Product} data type to represent product details. In the monolithic architecture, this shared model is used freely across modules, for example, in method parameters, return types, and internal logic for validating orders or updating stock}.

After decomposition, the \textit{InventoryManagement} microservice becomes the owner of product-related data, as it is responsible for managing stock and product attributes. However, the \textit{OrderManagement} microservice still directly depends on the \textit{Product} type, for example, in its method parameters, return types, or internal logic when creating or validating orders. 

This dependency introduces several limitations:
\begin{itemize}
    \item Any change to the \textit{Product} type in \textit{InventoryManagement}, even one unrelated to order processing, can break functionality in \textit{OrderManagement}.
    \item \textit{OrderManagement} cannot evolve its order creation logic without being tightly coupled to the structure and semantics of the \textit{Product} type defined in another service.
    \item \textit{OrderManagement} cannot evolve its order creation logic without being tightly coupled to the structure and semantics of the \textit{Product} type defined in another service.
\end{itemize}

To enable microservice extraction and ensure autonomy, the shared \textit{Product} model must be replaced with a replicated, service-specific representation. This is achieved by having \textit{OrderManagement} maintain a local copy of the product data it needs.

This transformation allows \textit{OrderManagement} to operate independently, using its own internal representation of product data—decoupled from the source model in \textit{InventoryManagement}. 

\subsubsection*{Strategy} We must identify these data type dependencies and refactor them appropriately to achieve separation of concerns and enable independent evolution, so that we can separate the microservices smoothly.
This may include:
\begin{itemize}
    \item Centralizing ownership in a single microservice and treat the data type as belonging exclusively to the microservice where it was originally defined. 
    \item Replicating the data type across microservices if both services require local access to the data type. 
    \item Using a Proxy Microservice in cases where one service acts primarily as a consumer and does not own or modify the data, it can serve as a proxy. 
\end{itemize}

\subsubsection*{Benefits}
The main benefits of this refactoring include:
\begin{itemize}
    \item More cohesive microservices.
    \item Reduced coupling between services.    
\end{itemize}

\subsubsection*{Challenges}
However, one of the main challenges of this refactoring is:
\begin{itemize}
    \item Correctly identifying where the boundaries should be drawn, especially when data usage spans multiple contexts.
    \item Managing data fragmentation, which may increase the complexity of inter-service communication.
\end{itemize}

\subsubsection*{Mechanics}
\begin{enumerate}
    \item Identify where the data type is used (for example, as attribute types in classes, as parameter or return types in methods, as method invocations tied to the data type).
    \item Choose a refactoring strategy. There are three main ways of doing this:
    \begin{enumerate}
        \item Assuming it belongs only to the microservice where it was first defined:
        \begin{enumerate}
            \item Method invocations:
            \begin{enumerate}
                \item Create an interface with the same name as the data type that defines the required operations on the data type.
                \item Implement this interface in a service that communicates with the owning microservice.
                \item Change method invocations from local calls to calls to the service that owns the data types and its methods, using the refactoring \textbf{``Replace Method Call with Service Call''} (Section~\ref{changeCalldoc}).
            \end{enumerate}
            \item Attributes, parameters, and return types:
                \begin{enumerate}
                    \item Replace direct usage of the shared type with a Data Transfer Object (DTO), that will represent that data type in the microservice and that will be sent through the service calls.
                    \item Use the refactoring \textbf{``Create Data Transfer Object''} (Section~\ref{dto_doc}).
                \end{enumerate}
                \item Modify the consuming service to use the DTO and interface instead of the original shared type.
        \end{enumerate}
        \item Keep it in both microservices if both services require local access to the data type:
        \begin{enumerate}
            \item Replicate the data type in both microservices.
            \item Use event sourcing or data replication to keep the copies in sync. Check the refactorings  \textbf{``Replace Method Call with Service Call: asynchronous''} (Section~\ref{changeCalldoc}) and \textbf{``Replicate Data Across Microservices''} (Section~\ref{replicate_data_doc}).

        \end{enumerate}
        \item Keep it in both microservices, but one of them is a proxy, if one service only consumes the data.
        \begin{enumerate}
            \item Introduce a proxy microservice that exposes the required operations.
            \item The proxy delegates requests to the owning service, abstracting the dependency.
        \end{enumerate}

    \end{enumerate}
\end{enumerate}
\subsubsection*{Example of application}
This example focuses on the centralized ownership strategy, assuming the \textit{Product} data type belongs exclusively to the \textit{InventoryManagement} microservice.

In the monolithic system, the \textit{OrderService} in \textit{OrderManagement} directly depends on the \textit{Product} type, using it as an attribute and invoking methods on it, which can be seen in 
Listing~\ref{order_before_doc}.

\begin{lstlisting}[language=java, label=order_before_doc, caption=\textit{OrderManagement} microservice before the refactoring.]
// Candidate for the OrderManagement microservice
public class OrderService {
    private ProductService productService;
    public OrderService(ProductService productService) {
        this.productService = productService;
    }
    public void createOrder(Order order) {
        // Perform order creation logic

        // Directly access the ProductService to get product information
        Product product = productService.getProductById(order.getProductId());
        // Use the product to complete the order creation process
    }
}

\end{lstlisting}

To resolve it, we create a \textit{ProductDTO} to use for transferring the \textit{Product} data between the microservices through service calls, and we modify the return types, attributes and parameters in the service's communications to use the DTO.
We, then, create a \textit{ProductInterface} that defines the necessary methods invocations to interact with \textit{Product} data in the \textit{InventoryManagament} microservice. The \textit{ProductService} implements this interface and makes the requests to the \textit{InventoryManagament} microservice that owns the data type \textit{Product}. 

This way, we have to replace the local method invocations in the \textit{Order} service that involves the \textit{Product} data type with calls to the \textit{ProductService} interface, which will make service calls to the \textit{InventoryManagament} microservice to retrieve or manipulate the \textit{Product} data. 

Lastly, we update the \textit{Order} service to use the new data type and the \textit{ProductService} interface for method invocations. All changes performed to the \textit{Inventory} microservice can be seen in Listing~\ref{inventory_break_doc} and all changes performed to the \textit{OrderManagement} microservice can be seen in Listing~\ref{order_break_doc}.

\begin{lstlisting}[language=java, label=inventory_break_doc, caption= \textit{InventoryManagement} microservice exposes product data via a DTO.]
// Candidate for the InventoryManagement microservice
@Service
public class InventoryService {
    public ProductDto getProductById(Long productId) {
        // Logic to fetch product data from inventory or other source
        // ...
        // Assume 'product' holds retrieved product data
        ProductDto product = new ProductDto();
        product.setId(productId);
        product.setName("Example Product");
        product.setPrice(BigDecimal.valueOf(9.99));
        return product;
    }
}
public class ProductDto {
    private Long id;
    private String name;
    private BigDecimal price;
    // Getters and setters
}

\end{lstlisting}

\begin{lstlisting}[language=java, label=order_break_doc, caption={\textit{OrderManagement} microservice after the refactoring, using a DTO and interface to decouple from the \textit{Product} type.}]
// Candidate for the OrderManagement microservice
public class ProductDto {
    private Long id;
    private String name;
    private BigDecimal price;
    // Getters and setters
}
public interface ProductInterface {
    ProductDto getProductById(Long productId);
}

@Service
public class ProductService implements ProductInterface {
    private final RestTemplate restTemplate; // or any HTTP client

    public ProductService(RestTemplate restTemplate) {
        this.restTemplate = restTemplate;
    }

    public ProductDto getProductById(Long productId) {
    // Make an HTTP request to the InventoryService to fetch the product
        String inventoryServiceUrl = "http://inventory-service/api/products/" + productId;
        ResponseEntity<ProductDto> response = restTemplate.getForEntity(inventoryServiceUrl, ProductDto.class);
        return response.getBody();
    }
}

@Service
public class OrderService {
    private final ProductService productService;

    public OrderService(ProductService productService) {
        this.productService = productService;
    }

    public void createOrder(OrderDto orderDto) {
        // Process the order details
        // Retrieve product information from the ProductService
        Long productId = orderDto.getProductId();
        ProductDto product = productService.getProductById(productId);
        // Continue order processing using product data
    }
}
\end{lstlisting}

\subsection{Shared Code Isolation}
\subsubsection*{Context and Motivation}
During the process of extracting microservices from a monolithic system, it is common to find shared code artifacts, such as utility classes, interfaces, or abstract classes, that are used across multiple components. In a monolith, these shared files are typically accessed through direct references, benefiting from a unified codebase and runtime environment. 

However, once services are separated, this shared usage becomes problematic. Sharing code across microservices can create tight coupling, making it harder for each service to operate and evolve independently. When multiple services rely on the same file, even a small change can ripple through the system, affecting deployment workflows, version control, and fault isolation. It also restricts the autonomy of individual services, particularly when the shared code contains business logic or is subject to frequent updates. To support independent evolution and resilience, it becomes necessary to rethink how shared code is managed in a distributed architecture.

\subsubsection*{Example} Consider the above-mentioned system that manages order processing
and inventory tracking, which was initially implemented as a monolith. 
\begin{itemize}
    \item \textbf{Scenario 1}: Imagine it contains a file called \textit{Utils.java} that defines multiple functions useful for this domain, but doesn't handle any business logic, like date formatting and string manipulation. If the microservice \textit{OrderManagement} and the microservice \textit{InventoryManagement} both use these functions from that file, as the system transitions to microservices, both domains are extracted into separate services. However, they still rely on \textit{Utils.java}, which exists only in the original monolith. This shared dependency creates a barrier to full service independence and complicates deployment.
    \item \textbf{Scenario 2}: Consider that multiple components rely on a shared module called \textit{ValidationLib}, which contains general purpose validation logic. This module is updated periodically to reflect new validation rules. As the system is decomposed into microservices, the \textit{OrderManagement} and \textit{InventoryManagement} services continue to depend on \textit{ValidationLib}. Because the code changes over time and consistency is important, the shared dependency introduces coordination overhead and risks of version drift, signaling the need for a refactoring strategy that supports reuse without tight coupling.
    \item \textbf{Scenario 3}: Imagine we have a component called \textit{PricingCalculator} that is responsible for applying business rules to compute discounts and taxes. This logic is used by both the \textit{Order} and \textit{Billing} domains. As these domains are extracted into separate microservices, they still require access to the pricing logic. However, the logic is complex, frequently updated, and critical to business operations. Keeping it as a shared file or duplicating it would lead to inconsistencies and maintenance challenges, making it clear that refactoring is needed to centralize and expose this logic in a more modular way.
\end{itemize}

\subsubsection*{Strategy} The strategy to support independent service evolution when we have shared code depends on the nature of the dependency and stability of the code:
\begin{itemize}
    \item \textbf{Scenario 1:} If we have stable utility code without business logic, the code can be safely duplicated across microservices. By duplicating such files, each microservice maintains autonomy and avoids runtime dependencies on external modules. Although this approach introduces code duplication, the trade-off is justified by the gain in modularity and resilience. It is important to ensure that duplicated files are well documented and versioned to reduce the risk of divergence over time.
    \item \textbf{Scenario 2:} If the code is unstable or frequently changing, we should extract it into a shared library that is versioned and centrally maintained, which enables reuse while controlling updates.
    \item \textbf{Scenario 3:} If it contains shared business logic, then it is probably best to encapsulate it in a dedicated microservice that exposes its functionality via an API. This ensures consistency and avoids duplication, while supporting independent deployment and scaling.
\end{itemize}

\subsubsection*{Benefits} Some of the benefits of this refactoring are:
\begin{itemize}
    \item Scenario 1: 
        \begin{itemize}
            \item Service autonomy through code ownership, each microservice can evolve without being constrained by shared dependencies.
            \item Reduced coupling from eliminating shared artifacts: reduces build time and runtime dependencies.
            \item Simplified deployment and fault isolation, as services no longer rely on a common module failures caused by changes in shared code are avoided.
        \end{itemize}
    \item Scenario 2:
        \begin{itemize}
            \item Centralized code management.
            \item Easier to maintain consistency.
        \end{itemize}
        
    \item Scenario 3:
    \begin{itemize}
        \item Single source of truth for business logic.
        \item Promotes consistency across services.
        \item Enables independent scaling and versioning of shared logic.
    \end{itemize}
\end{itemize}

\subsubsection*{Challenges} The main challenges of this refactoring are:
\begin{itemize}
    \item Scenario 1:
    \begin{itemize}
        \item Manual synchronization is required, if changes are made to the duplicated file, these changes will need to be manually replicated across all microservices that use the file.
        \item Risk of inconsistent behaviour.
        \item Difficult traceability and version control, as code duplication can create inconsistencies.
    \end{itemize}
    \item Scenario 2:
    \begin{itemize}
        \item Tighten build time coupling.
        \item Requires coordinated releases.
        \item Limits tech stack flexibility.
    \end{itemize}
    \item Scenario 3:
    \begin{itemize}
        \item Adds network latency.
        \item Requires robust fault tolerance.
        \item Increases infrastructure complexity.
    \end{itemize}
\end{itemize}

\subsubsection*{Mechanics}
\begin{itemize}
    \item \textbf{Scenario 1:}
    \begin{enumerate}
        \item Identify the utility classes or functions that are used across services.
        \item Confirm that they do not contain any business logic or domain specific, and therefore, don't fall into other scenarios.
        \item Copy the file into each service codebase.
    \end{enumerate}
        \item\textbf{Scenario 2:}
        \begin{enumerate}
            \item Extract shared logic into a standalone module or package.
            \item Publish the library to a private package registry (e.g. Maven, npm, Docker, etc.)
            \item Update each service to depend on the library.
            \item Establish a release and update process to manage changes.
        \end{enumerate}
        \item \textbf{Scenario 3:}
            \begin{enumerate}
                \item Extract the shared logic into a new microservice.
                \item Define a clear API contract (RESTfull HTTP, gRPC, etc.).
                \item Implement client-side integration to consume the service.
            \end{enumerate}
            Note: This is very similar to the mechanics of \textbf{``Replace Method Call with Service Call''} (Section~\ref{changeCalldoc}), which is in its simplified version here.
\end{itemize}

\subsubsection*{Example of application}
\begin{itemize}
    \item \textbf{Scenario 1:} We begin by confirming that \textit{Utils.java} only contains functions without business logic. Then copy \textit{Utils.java} into the codebase of both \textit{OrderManagement} service and \textit{InventoryManagement} service.
    
    \item\textbf{Scenario 2:} We identify the shared validation logic used across multiple services and extract it into a standalone module named \textit{ValidationLib}. Publish this module to a private Maven registry using semantic versioning to manage updates. Each microservice, \textit{OrderManagement} and \textit{InventoryManagement}, should update its build configuration to include \textit{ValidationLib} as a dependency.

    \item \textbf{Scenario 3:} Isolate the discount calculation logic and migrate it to a dedicated microservice called \textit{DiscountService}. Define a clear API contract, such as a RESTful endpoint \textit{/calculate-discount}, to expose the required functionality. Integrate both \textit{OrderManagement} and \textit{BillingManagement} services with \textit{DiscountService} via HTTP calls. Finally, remove the original shared discount logic from both services to eliminate redundancy. Note: The communication strategy between services can be either synchronous or asynchronous, depending on system requirements, this aligns with the approach described in the mechanics of \textbf{``Replace Method Call with Service Call''} (Section~\ref{changeCalldoc}).
\end{itemize}

\section{Related work}\label{related_work}
This section presents some refactoring publications related to our study. We cover a few different perspectives, including foundational works on the concept of refactoring, studies that present refactoring catalogs applied in different contexts, and those specifically focused on microservices. While many \textit{patterns} have been written to support designing microservices and cloud-native systems~\cite{Sousa2018c,Sousa2015,sousa2016engineering,Maia2022,maia2024patterns,maia2025container,Sousa2017,Sousa2018b,Albuquerque-2022,Albuquerque-2023,Albuquerque-2024,Albuquerque-2025,Dobaj2019,cdp_aac,zimmermann2022patterns,cloudpatterns2021brown}, fewer works delve into how to migrate to such architectures.

One of the main materials on refactoring is the book \textit{``Refactoring: Improving the Design of Existing Code''} by Fowler and Beck~\cite{Fowler99}. The authors introduce the principles and best practices of refactoring, guiding developers on when and where to start analyzing code for improvements. However, one of the main contributions of the book is its comprehensive catalogue of refactorings, which addresses aspects such as code readability, class and object structure, modularization, and data processing. In our study, we also present a catalog of refactorings, but our focus is on supporting the transition from monoliths to microservices, specifically presenting refactorings related to handling dependencies.

Other works have also tried mapping refactorings for specific contexts, as the following paragraphs briefly show. 

Rizvi and Khanam~\cite{Rizvi2011} explore the combination of Aspect-Oriented Programming (AOP) and refactoring as a strategy to handle the continuous evolution of software. They propose a catalogue of refactorings that enables the extraction of crosscutting concerns from legacy procedural code, specifically in C, using AOP concepts, to make the code more understandable, modular, and easier to maintain. The proposed catalogue contains 10 aspect-oriented refactorings for procedural code. Similar to our work, the authors present a refactoring catalog to address software evolution, aiming to improve modularity and maintainability. Both works focus on code-level refactorings, although our approach specifically targets the transformation process toward a microservices architecture.

Oberlehner et. al.~\cite{Oberlehner2021} propose a catalogue of refactoring operations specific to systems based on the IEC 61499 standard, which is widely used in the development of industrial automation systems and cyber-physical systems. Their goal is to improve the quality of these systems, making them more understandable, maintainable, and modular. The proposed catalogue contains six refactoring groups. The domain addressed by the authors is different from ours, but both works propose refactorings addressing internal parts of the system components, aiming for gradual improvements.

Two articles led by Stocker \cite{Stocker2023,Stocker2024} present a catalog of 15 refactorings focused on APIs and their architectural elements. Currently, the full Interface Refactoring Catalog (IRC) consists of 24 refactorings. The authors explain that these works are motivated by the challenges encountered in the evolution of distributed systems based on remote APIs. While internal code refactoring is already a well-established practice, API refactoring still lacks structured guidelines. Similar to our work, these works focus on architectural concerns. Our work aims to support the transition to microservices, which may include API changes as part of broader transformations.

Isaenko~\cite{Isaenko2018} presents a catalogue of eight refactorings for microservices-based systems, helping to address software degradation and compensate for technical debt. The goal is to deal with the challenges of refactoring these types of systems, which are both distributed and complex. Thus, the work identifies and documents patterns that help developers evolve and maintain microservices efficiently, enabling changes to be made safely and aligning with good architectural practices. This work pursues a similar goal to ours,  but differs in granularity and application context. Isaenko’s catalog provides broader strategies intended to support the ongoing maintenance of microservices-based systems. In contrast, our catalog emphasizes fine-grained refactorings applied during the migration from monolithic systems to microservices.

Another work directly related to microservices is that of Tighilt et. al.~\cite{Tighilt2020}, who presents a catalog of 16 microservice antipatterns, organized into the categories of design, implementation, deployment, and monitoring. Each antipattern is described, with its implementation and possible refactoring solutions to mitigate it. The authors highlight that the results can be useful by helping practitioners identify and prevent inappropriate practices in microservices development. Although both our work and Tighilt et al.'s share the same motivation of supporting the transformation from monolithic to microservices architectures, they differ in focus, granularity, and applicability. While both works aim to improve system quality and maintainability, Tighilt et al.'s focus is on preventing design failures, and our work emphasizes practical refactorings for architectural evolution during migration.

The catalogue of refactorings presented in our article evolved from the catalogue of refactorings proposed by Pinto~\cite{pinto_refactoring_2019}. The main objective of our work is to systematize existing knowledge of how to migrate from monoliths to microservices, as a catalogue of refactorings that can mitigate common difficulties. Our work refines and expands the initial catalog proposed by Pinto by rethinking the refactorings, incorporating examples and context, and introducing additional refactorings derived from a literature review and empirical study.

\section{Conclusion}

Migrating monolithic systems to microservices architectures is a challenging process that requires systematic methodologies. This article contributes a comprehensive catalogue of refactorings specifically designed to \textit{preparing} dependencies to make a future service extraction easy. Many approaches have been proposed for defining service boundaries, our work tries, instead, to provide actionable, code-level refactorings that enable developers to incrementally prepare the ground for service extraction.

There is ample opportunity for further refinement and expansion of the catalogue. Future iterations can incorporate additional edge cases, more diverse examples, and new refactorings to address challenges that may arise during migration. We envision this catalogue as a living resource, enriched by contributions from other researchers and practitioners, ensuring its continued relevance and utility.

By focusing on refactoring dependencies and preparing for service extraction, this work provides a practical and systematic guide for developers undertaking the migration from monolithic systems to microservices architectures. It lays the foundation for advancing both the methodology and tooling required to streamline this complex transformation process, ultimately empowering developers to achieve successful and sustainable migrations. Ultimately, the availability of tools and ability to automate service extraction, may contribute to an easier adoption of microservices and reduce the premium to the projects' cost and risks usually associated with transitioning to microservices~\cite{fowler_microservicespremium}.

\begin{credits}
\subsubsection{\ackname} We would like to thank our EuroPLoP shepherd---Valentino Vranić---and all the members of our EuroPLoP workshop---Dionysis Athanasopoulos, João Daniel, Pierre Schnizer, Stefan Kapferer and Tammo van Lessen. Their comments and insights allowed us to gain perspective on our own work and to be able to improve it considerably. 

This project has received funding from the European Union's Horizon Europe research and innovation programme under the Grant Agreement 101081020.

\subsubsection{\discintname}
The authors have no competing interests to declare that are relevant to the content of this article.
\end{credits}
\bibliographystyle{splncs04}
\bibliography{references}

\end{document}